\documentclass{moriond}

\def\be{\begin{equation}}
\def\ee{\end{equation}}
\def\bea{\begin{eqnarray}}
\def\eea{\end{eqnarray}}

\newcommand{\Photo}{}

\begin{document}

\vspace*{-1.5cm}

\begin{flushright}
CERN-TH-2016-115
\end{flushright}

\vspace*{4cm}

\title{TOWARDS THE COMPRESSION OF PARTON DENSITIES THROUGH MACHINE LEARNING ALGORITHMS}

\author{Stefano Carrazza$^{1*}$ and Jos\'e I. Latorre$^2$}

\address{$^{1}$Theoretical Physics Department, CERN, Geneva,
  Switzerland, $^*$Speaker\\ $^2$Departament d'Estructura i Contituents de
  la Mat\`eria, Universitat de Barcelona }

\maketitle\abstracts{ One of the most fascinating challenges in the
  context of parton density function (PDF) is the determination of the
  best combined PDF uncertainty from individual PDF sets. Since 2014
  multiple methodologies have been developed to achieve this goal. In
  this proceedings we first summarize the strategy adopted by the
  PDF4LHC15 recommendation and then, we discuss about a new approach to
  Monte Carlo PDF compression based on clustering through machine
  learning algorithms.}

\paragraph{The PDF4LHC15 recommendation and tools for LHC Run II}

In October 2015 the PDF4LHC Working Group released a new set of
guidelines for the combination of PDF sets, known as the ``PDF4LHC15
recommendation'' published in Ref.~\cite{Butterworth:2015oua}. This
updated recommendation proposes the construction of a combined prior
PDF set of Monte Carlo (MC) replicas, where each replica comes from
global PDF determinations. The prior set is then compressed to a
minimal number of PDF members through reduction algorithms specialized
in the removal of information redundancy.

The PDF4LHC15 prior consists in $N_{\rm rep}=900$ MC replicas from
NNPDF3.0~\cite{Ball:2014uwa}, CT14~\cite{Dulat:2015mca} and
MMHT2014~\cite{Harland-Lang:2014zoa}. Eigenvectors from CT14 and
MMHT2014 are transformed into MC replicas through the method developed
by Watt and Thorne in Ref.~\cite{Watt:2012tq} and implemented in the
{\tt LHAPDF6}~\cite{Buckley:2014ana} library. The PDF sets entering in
the current combination satisfy requirements which guarantee the
consistency of results: use global datasets, compute theoretical
predictions and DGLAP in the GM-VFNS, set $\alpha_s$ to the PDG
average~\cite{Agashe:2014kda}.

The subsequent step consists in removing the redundant information
from the prior set through reduction algorithms. For the PDF4LHC15
recommendation we have used 3 different strategies:
CMC-PDF~\cite{Carrazza:2015hva}, MC2H~\cite{Carrazza:2015aoa} and
Meta-PDF~\cite{Gao:2013bia}. The CMC-PDF approach outputs a subset of
MC replicas which preserves the statistical properties of the prior
set. The MC2H strategy provides a symmetric Hessian PDF set obtained
by using the MC replicas themselves as the basis of the linear
representation in combination with principal component analysis (PCA)
to reproduce the PDF covariance matrix with arbitrary precision. The
Meta-PDF approach refit each MC replica with a flexible
meta-parametrization, from which the best constrained combination are
found by diagonalization of the covariance matrix on the PDF space.

The delivery of results in the MC representation is useful when
considering regions where predictions are non-Gaussian, such as
searches at high-masses and generally wherever the PDF is probed at
large $x$. On the other hand, Hessian sets are useful for many
experimental needs, {\it e.g.} when using nuisance parameters, or when
high accuracy is required.
The PDF4LHC15 recommendation delivers sets at NLO and NNLO with
$n_f=4,5$, noted as: {\tt PDF4LHC15\_mc}, the compressed Monte Carlo set
with $N_{\rm rep}=100$ obtained with CMC-PDF; {\tt PDF4LHC15\_100}, the
symmetric Hessian set with $N_{\rm eig}=100$ obtained with MC2H; {\tt
  PDF4LHC15\_30}, the symmetric Hessian set with $N_{\rm eig}=30$
obtained with Meta-PDF.

Finally, thanks to developments for the PDF4LHC15 recommendation, a
recent Hessian reduction algorithm called Specialized Minimal PDF
(SMPDF) was published in Ref.~\cite{Carrazza:2016htc}. The SMPDF
methodology constructs PDFs designed to provide an accurate
representation of PDF uncertainties for specific processes or classes
of processes with a minimal number of PDF error sets.

\begin{figure}[t]
  \begin{centering}
    \includegraphics[scale=0.28]{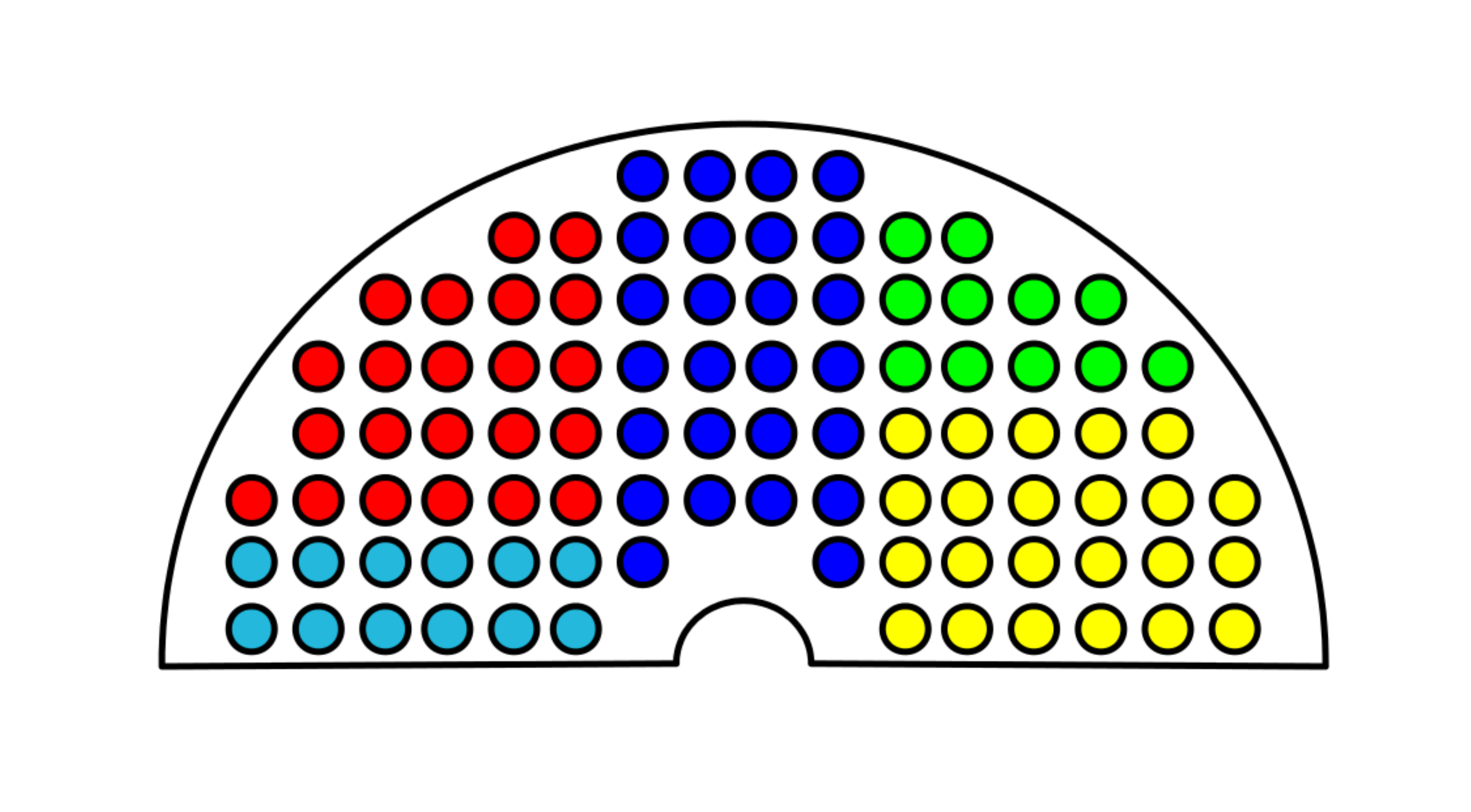}\includegraphics[scale=0.28]{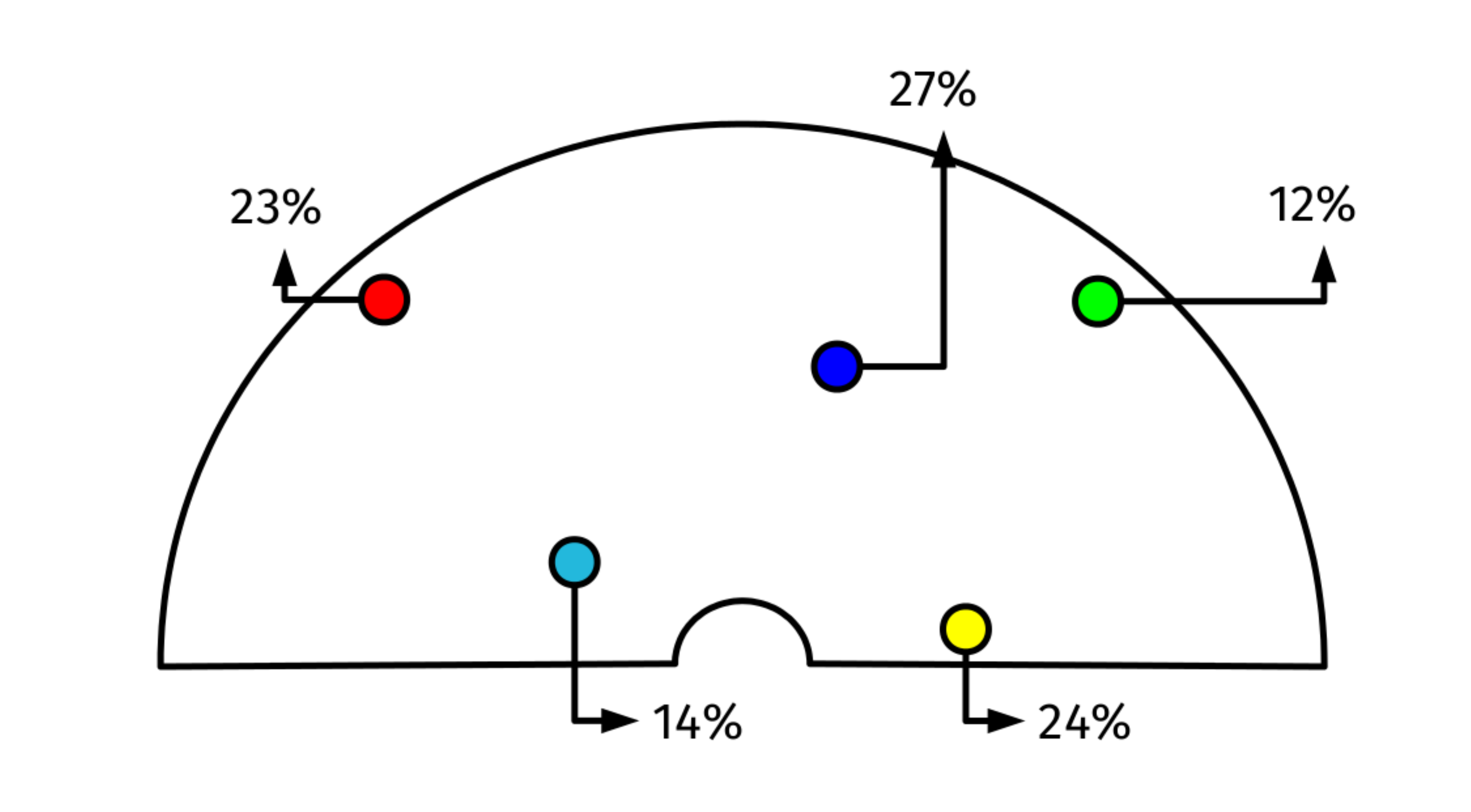}
    \par\end{centering}
    \caption{\label{fig:parliament} Illustrative example of the
      clustering idea. A parliament is composed by several elements
      (left), for each of them it is possible to determine the most
      representative exemplars and the fraction of elements similar to
      them (right).}
\end{figure}

\paragraph{Monte Carlo PDF compression through machine learning}

In the next paragraphs we present a new concept of compression
approach for MC PDFs based on clustering through machine learning
algorithms.

Let us consider a generic PDF set composed by a large number of MC
replicas. Starting from a simple visual inspection we observe that
groups of replicas have similar shapes, positions and lengths. This
observation suggests that in the PDF space there is a limited number
of shapes and directions privileged by replicas and so, if we want to
reduce the number of members contained in a PDF set we should extract
the most important replicas and their respective weights.
This observation is illustrated in Figure~\ref{fig:parliament} by the
analogy of the politicians and their parties in a parliament which in
our case study are identified to PDF replicas. The left plot shows the
initial distribution of elements. In this example similar objects are
identified by a color, and we have 5 groups. The right plot shows the
most representative exemplars of each group and their weight. The next
step is to setup a clustering algorithm able to identify the number of
groups, the most representative exemplars and their weights.

Here we use Affinity Propagation (AP), by messaging passing algorithm
presented in Ref.~\cite{Frey} where the authors show its impressive
capability of grouping data with complex structure. The choice of this
particular algorithm is motivated by its capability of determining
automatically the number of final clusters and its members without
requiring as input an {\it a prior} knowledge or guess of the number
of clusters. The only requirement of AP is to set a distance
definition to quantify the similarity between elements of a given
ensemble of PDF replicas. In the AP approach, we construct a
similarity matrix, defined as:
\begin{equation}
  S_{i,j} = -d(\ell_i,\ell_j),
\end{equation}
where $d(\ell_i,\ell_j)$ is the distance estimator defined by the user. We
performed the current analysis with the squared euclidean distance
between the arc-length of replicas defined as:
\begin{equation}
  \ell_k=\sum_{\alpha=-n_f}^{n_f} \int_0^1
  \sqrt{1+\left(\frac{df^{(k)}_\alpha(x,Q)}{dx}\right)^2} dx,
\end{equation}
where $k$ is the replica index and $\alpha$ runs over the $n_f$
independent PDF flavors at the factorization scale $Q$. We observed
that similar results are also obtained when using just the spatial
euclidean distance between replicas.

\begin{figure}[p]
  \begin{centering}
    \includegraphics[scale=0.55]{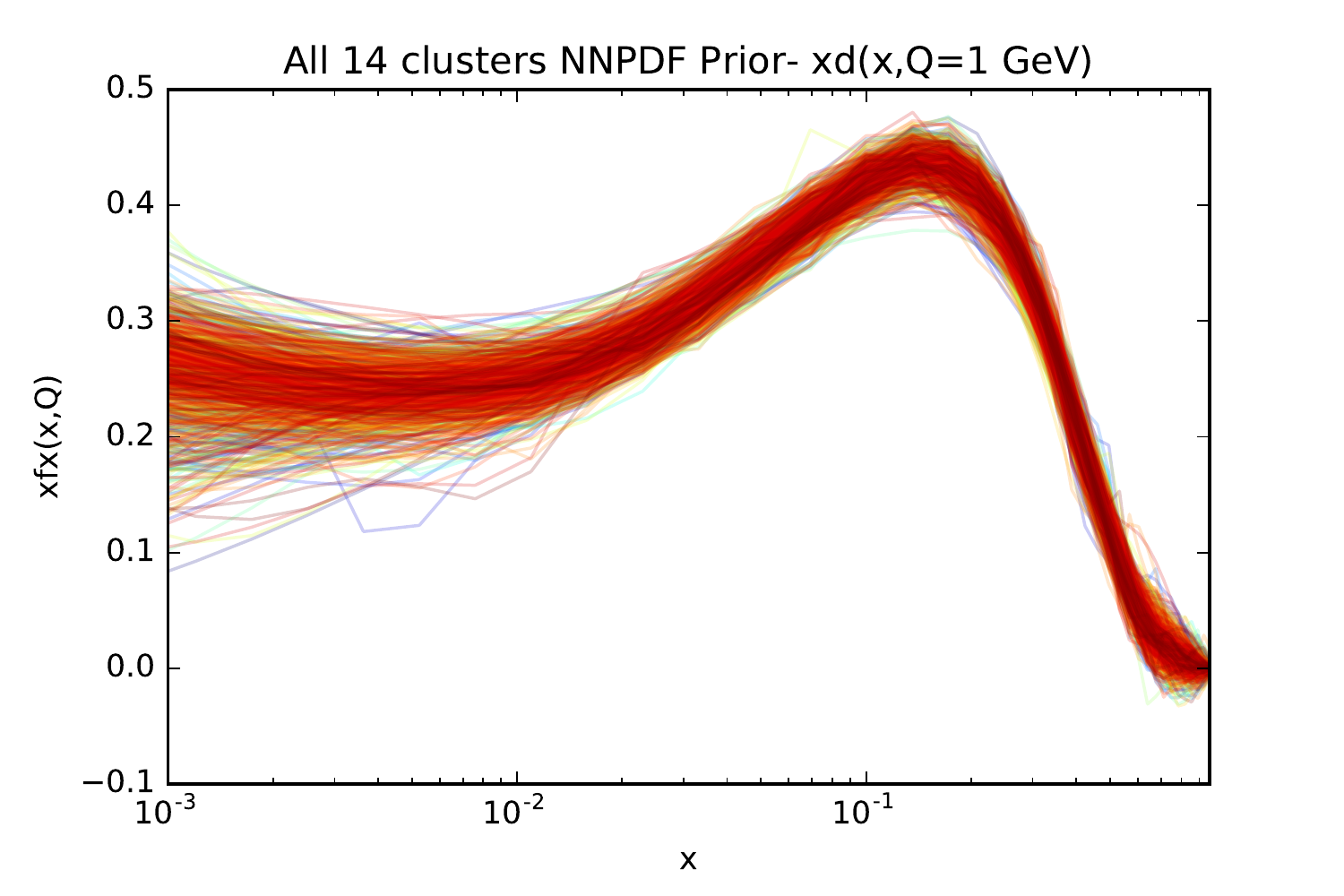}\includegraphics[scale=0.55]{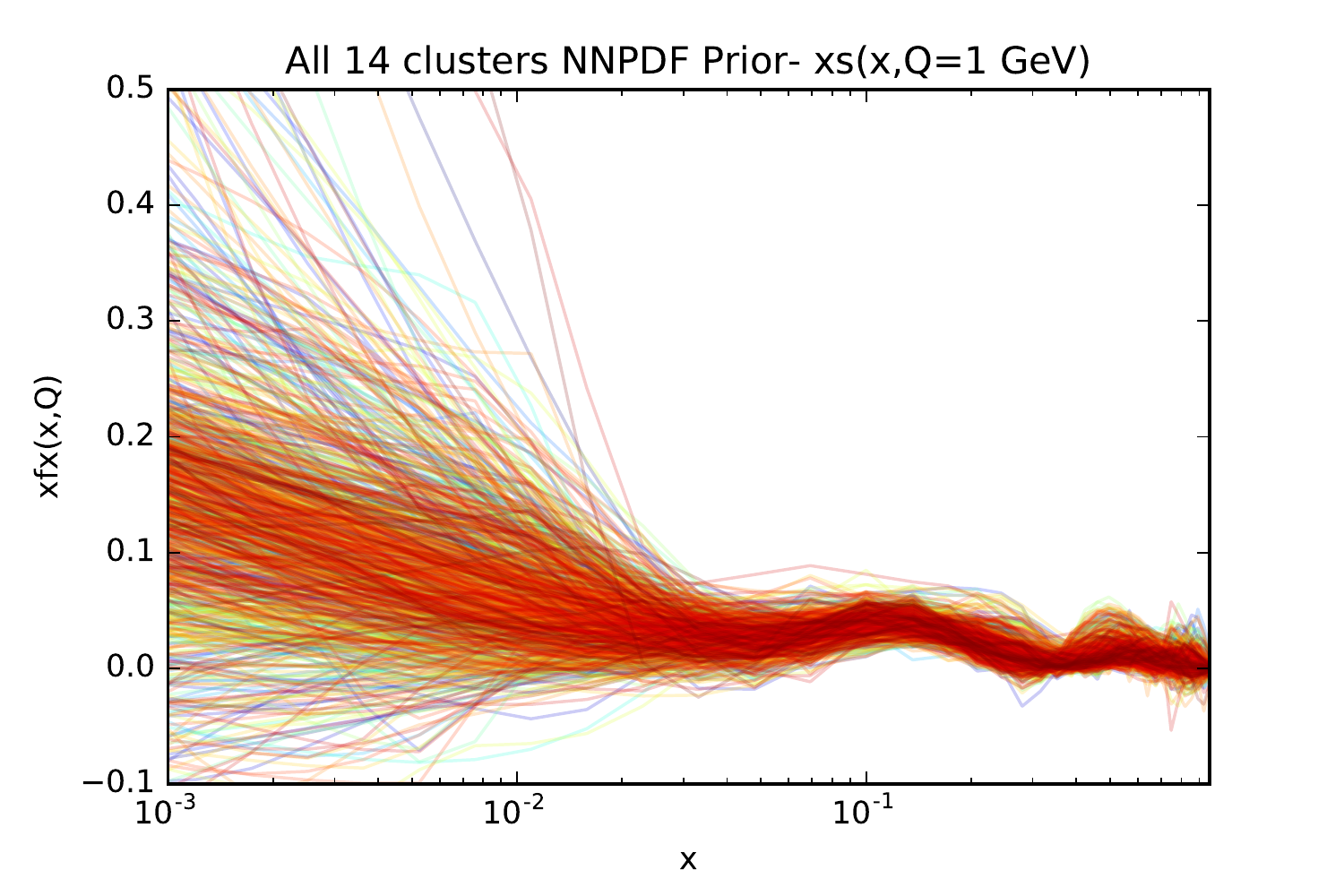}
    \par\end{centering}
    \caption{\label{fig:cluster} Examples of clustering of MC replicas
      using affinity propagation and arc-length distance metrics.}
\end{figure}

In Figure~\ref{fig:cluster} we show the results of this clustering
procedure for the NNPDF3.0 NLO set with $N_{\rm rep}=1000$
replicas. The AP algorithm identifies 14 clusters which are
represented by different colors for the down (left plot) and strange
(right plot) PDFs. The final step consists in computing the weight
associated to each cluster center exemplar. For each cluster $i$ we
define its associated weight, $w_i$, as:
\begin{equation}
  w_i = N_i/N_{\rm rep},\quad\sum_i w_i = 1,
\end{equation}
where $N_i$ is the number of elements contained in the cluster
$i$. The output of this procedure is a MC set of PDFs with $N_{\rm
  rep}=14$ MC replicas and a list of $N_{\rm rep}$ weights.

\begin{figure}[p]
  \begin{centering}
    \includegraphics[scale=0.55]{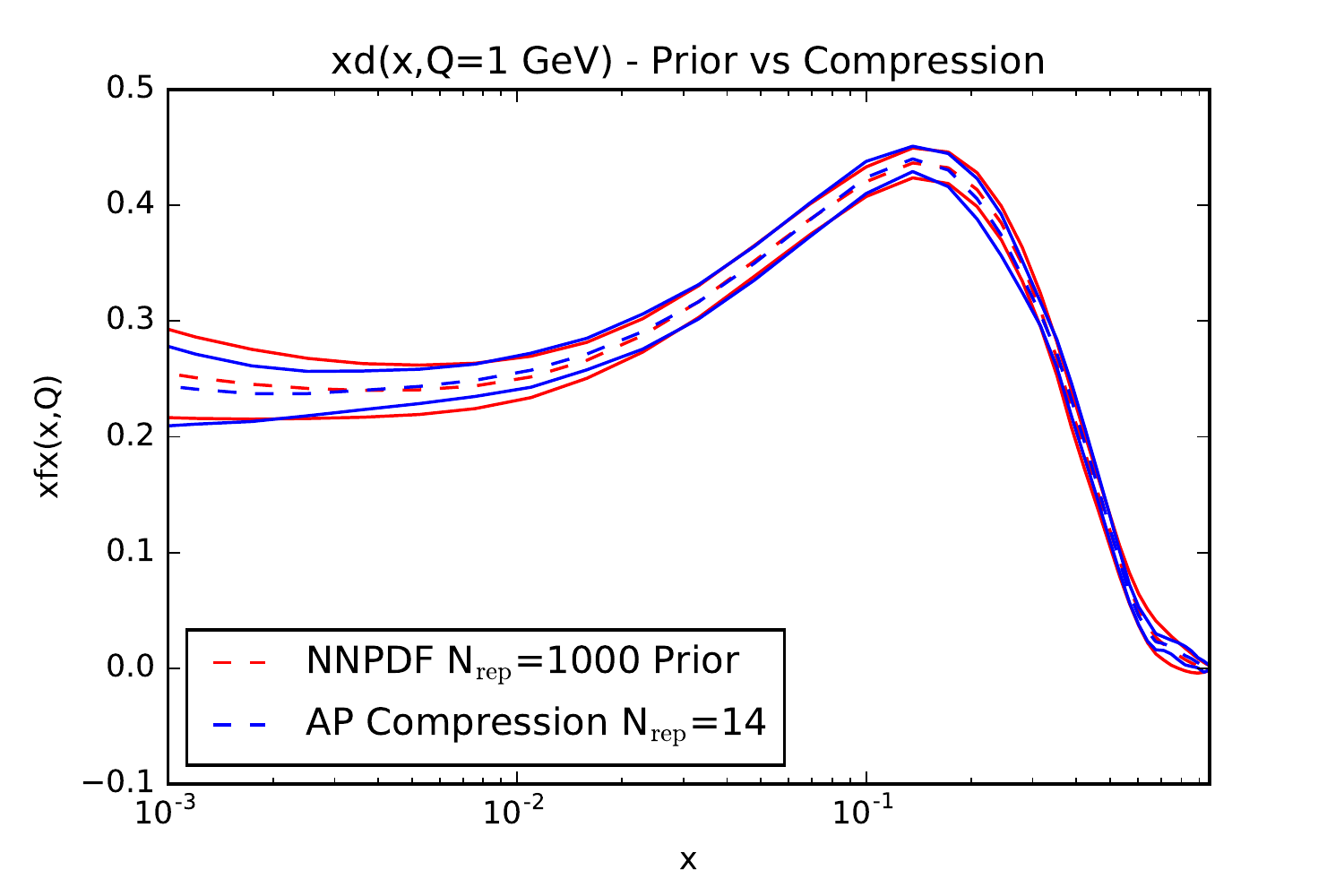}\includegraphics[scale=0.55]{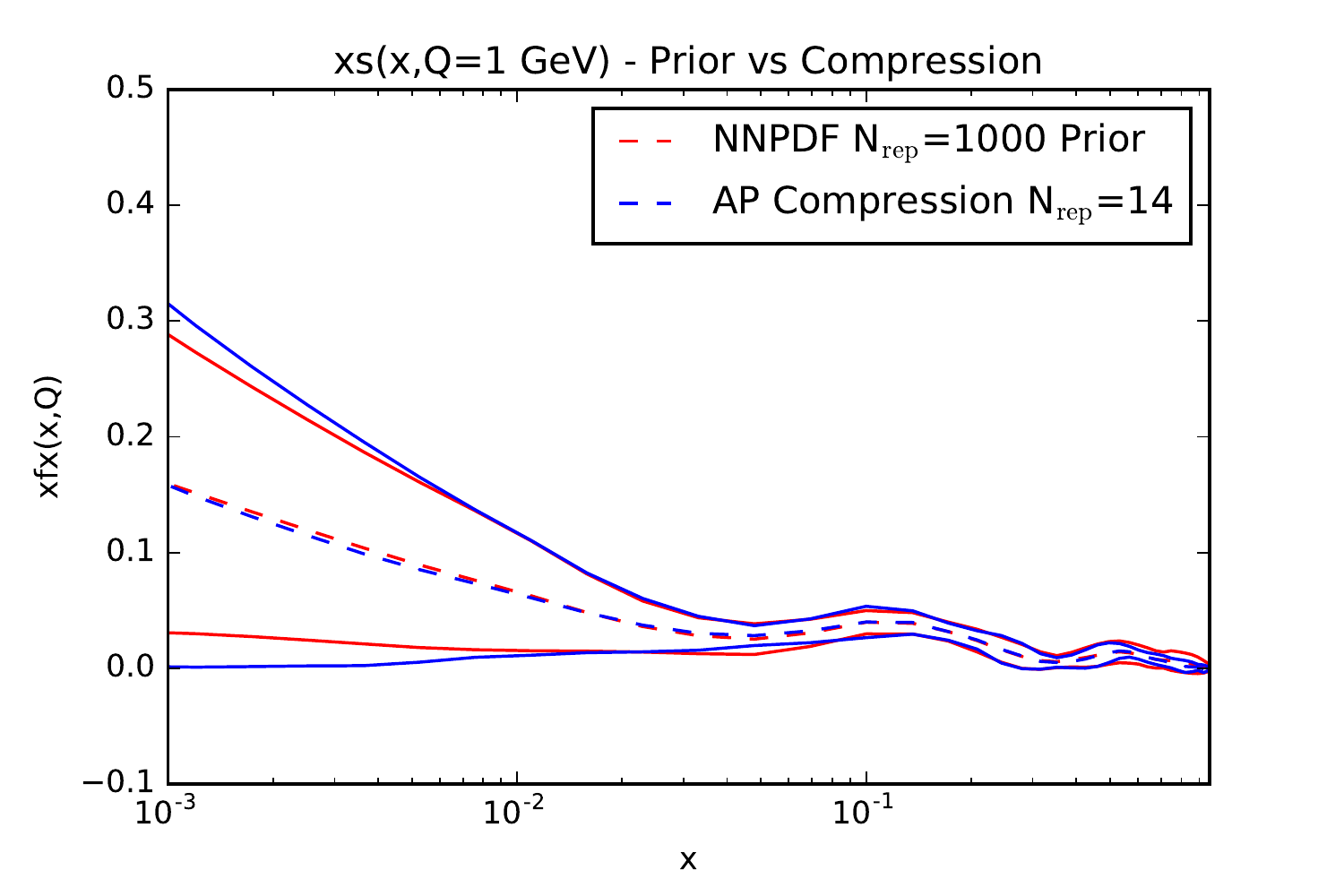}
    \par\end{centering}
    \caption{\label{fig:comparison} Comparisons for central values and
      uncertainties between the prior PDF set and the affinity
      propagation clustering compression.}
\end{figure}

\begin{figure}[p]
  \begin{centering}
    \includegraphics[scale=0.41]{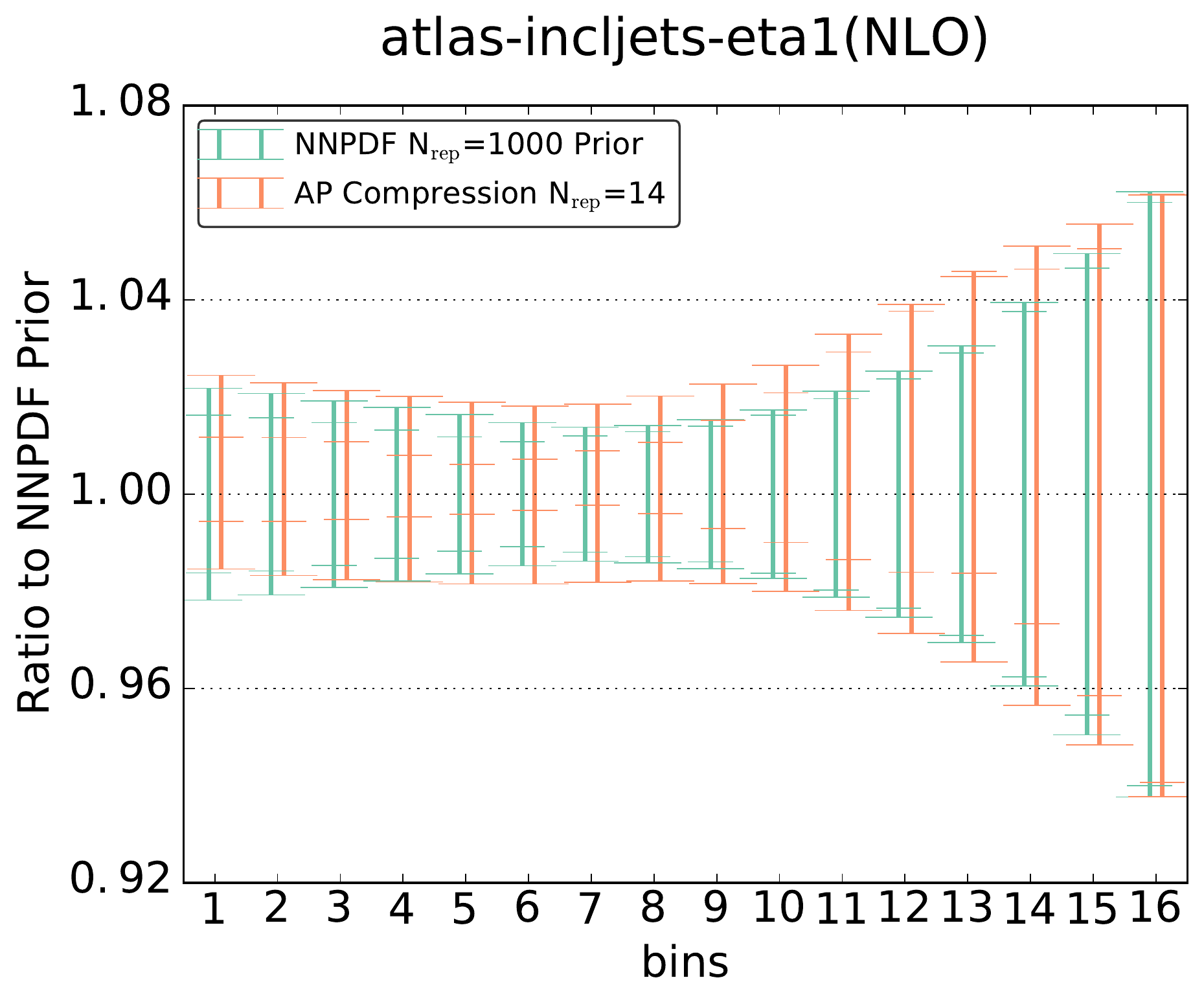}\includegraphics[scale=0.41]{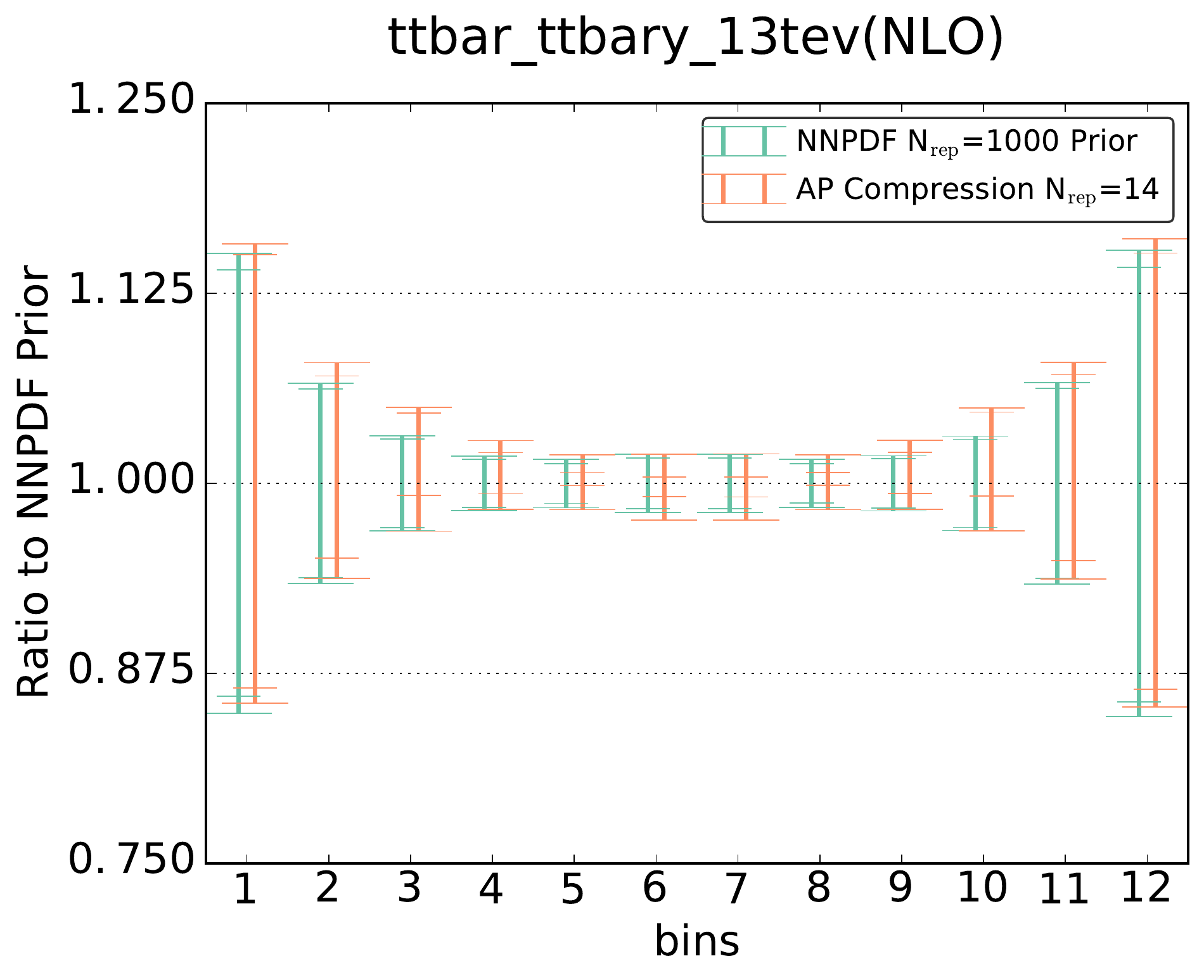}
    \par\end{centering}
    \caption{\label{fig:predictions} Comparisons of theoretical
      predictions for the prior PDF set and the affinity propagation
      clustering compression sets. Plots obtained with SMPDF.  }
\end{figure}

In Figure~\ref{fig:comparison} we compare the central value and its
uncertainty for the down and strange PDFs for the NNPDF prior and the
compressed set obtained with AP. For the AP PDF set we plot the
weighted mean and standard deviation. In general, we observe that a
good level of agreement is obtained. Furthermore, in
Figure~\ref{fig:predictions} we compare theoretical predictions of
both sets for the ATLAS inclusive jets setup with $|\eta|<0.3$ from
Ref.~\cite{Aad:2011fc} (left plot) and a $t\bar{t}$ rapidity
distribution at LHC with $\sqrt{s}=13$ TeV (right plot). Also in this
case, the level of agreement is satisfactory.

Similar results are obtained when using the PDF4LHC prior set. This
approach has two advantages in comparison to the CMC-PDF method: the
instantaneous computation time, and the possibility to compress to a
very lower number of replicas due to the flexibility of weights. We
are confident that this or similar approaches based on the idea of
weighting MC replicas are the right future direction to obtain fast
and outstanding compression performance of MC PDF sets.

\paragraph{Acknowledgments} S.C. is supported by the HICCUP ERC Consolidator grant (614577).

\end{document}